\documentclass[fleqn,twoside]{article}
\usepackage{espcrc2}
\usepackage{graphicx}
\usepackage[figuresright]{rotating}

\newcommand{\AmS}{{\protect\the\textfont2
  A\kern-.1667em\lower.5ex\hbox{M}\kern-.125emS}}
\title{A numerical study of a confined $Q\bar{Q}$ system in compact U(1) lattice gauge theory in 4D}
\author{M. Panero\address[dias]{Dublin Institute for Advanced Studies, 10 Burlington Road, Dublin 4, Ireland\\
        E-mail: panero@stp.dias.ie}
}

\begin{document}

\begin{abstract}
We present a numerical study about the confining regime of compact U(1) lattice gauge theory in 4D. To address the problem, we exploit the duality properties of the theory. The main features of this method are presented, and its possible advantages and limits with respect to alternative techniques are briefly discussed. In Monte Carlo simulations, we focus our attention onto the case when a pair of static external charges is present. Some results are shown, concerning different observables which are of interest in order to understand the confinement mechanism, like the profile of the electric field induced by the static charges, and the ratios between Polyakov loop correlation functions at different distances. \vspace{1pc}
\end{abstract}

\maketitle

\section{GENERAL SETTING}
Compact U(1) lattice gauge theory in 4D displays a confined phase, analogous to non-abelian gauge models; since different gauge theories may share the same (qualitative) mechanisms for confinement, it is interesting to study an external, static $Q\bar{Q}$ pair in this model, which provides a very simple confined system.

The fundamental d.o.f. of the pure gauge theory are $U_\mu(x)$ phase variables defined on the oriented bonds of an isotropic hypercubic lattice, the dynamics is described by Wilson action: $S = \beta \sum_{\mbox{\tiny{p}}} (1-\mbox{Re} U_{\mbox{\tiny{p}}} )$. For $ 0<\beta<\beta_c = 1.0111331(21) $ \cite{heplat0210010} the system is confined, while for $\beta> \beta_c$ it is in a deconfined (``Coulomb--like'') phase.

This theory enjoys a ``duality'' \cite{savit} property: via a group Fourier transform, the partition function and observable VEV's map to a dual formulation in terms of  ${^\star l}_{\mu}(x) \in \mathbf{Z}$ variables. In 4D the dual model is still a
gauge model of ``ferromagnetic'' nature:
\begin{equation}
\label{dualz}
Z = (2 \pi)^{4N} \prod \sum_{^\star l} e^{-\beta} I_{|d ^\star l|} (\beta)
\;
\end{equation}
(we follow notation of \cite{heplat9705019}, where the same method was used to study this model). This exact mapping allows one to get results for U(1) theory from simulations of the dual model. A $Q\bar{Q}$ pair (represented by Polyakov lines in the original model) can be introduced by means of a stack of topological defects $^\star n$ onto a set of plaquettes:
\begin{equation}
\label{dualzqq}
Z_{Q\bar{Q}} = \left( 2 \pi \right)^{4N} \prod \sum_{^\star l} e^{-\beta} I_{|d^\star l + ^\star n|}(\beta)
\;.
\end{equation}
Simulating the dual model gives some practical advantages from the
numerical point of view; the major improvement arises evaluating
ratios between Polyakov line correlators at increasingly large
interquark distance, which are affected by exponential
signal-to-noise ratio decay in direct simulations, whereas this
problem can be completely overcome in simulations of the dual
model --- see also \cite{heplat0007034}. This method was used for
$\mathbf{Z}_2$ gauge model in 3D
\cite{heplat9609041,heplat0211012,heplat0401032}, and it is a
possible alternative to other error reduction algorithms, like the
one proposed by L\"uscher and Weisz \cite{heplat0108014}, which
has been successfully used in a number of works about different
gauge theories
\cite{heplat0207003,heplat0209094,heplat0211038,heplat0306011,heplat0406008,heplat0406037},
including U(1) LGT in 4D \cite{heplat0309003,heplat0311016}.
However, this duality-inspired technique cannot be
straightforwardly generalized to non-abelian SU(N) theories.

\section{OBSERVABLES AND RESULTS}
We focused our attention onto the profile of the electric field longitudinal component induced in the symmetry plane between the static charges, onto the interquark potential, and force.

According to the dual superconductor picture, at large distance $\rho$ from the mid-point between the external charges, the electric field is expected to be described by a modified Bessel function:
\begin{equation}
\label{expectedex}
E_{x}(\rho) \propto m^2 K_0 \left( m\rho \right)
\;,
\end{equation}
$m$ being the mass of the dual gauge boson. This is based on a purely classical analysis; however, fluctuations of the flux tube can be included \cite{heplat9508017}: they induce logarithmic growth of the flux tube width \cite{heplat9510019}.

The interquark potential and force can be worked out from Polyakov loop correlators, which, according to the bosonic effective string scenario, are expected to behave as \cite{minami78,flensburgpeterson87}:
\begin{equation}
\label{expectedprpzero}
\langle P^\dagger (r) P(0) \rangle = \frac{e^{-\sigma rL -\mu L}}{ \left[ \eta \left(
i\frac{L}{2r} \right) \right]^{D-2} }
\;.
\end{equation}
Here $r$ is the interquark distance, $L$ is the Polyakov line length, $\sigma$ is the string tension, and $\eta$ denotes Dedekind's function. Correspondingly, the interquark potential $V(r)$ reads \cite{lueschersymanzikweisz80}:
\begin{equation}
\label{potential}
V(r) \simeq \sigma r + \mu -\frac{\pi(D-2)}{24r}\left( 1 + \frac{b}{r} \right) + \dots
\;,
\end{equation}
where we also included a possible ``boundary term'' contribution (depending on $b$) which was suggested in \cite{heplat0207003}, although it would break the expected open--closed string duality \cite{hepth0406205}. However, this picture is under debate: in various gauge models it was observed that, despite the fact that the string behaviour onset already appears at short distances (corresponding to $0.5 \sim 1.0$ fm or so), in that very same region the excited state spectrum does not match the expected effective string pattern \cite{heplat0401032,heplat0211038,heplat0309180,heplat0312019}.

\begin{figure}[htb]
\vspace{9pt}
\includegraphics[scale=0.44]{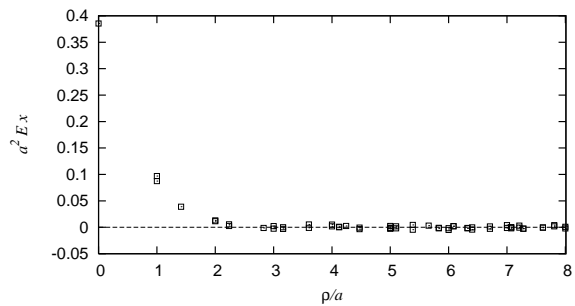}
\caption{$E_x$ in the symmetry plane between the external charges, as a function of $\rho$, the distance from the $Q\bar{Q}$ mid-point. Results for $\beta=0.96$, $d_{Q\bar{Q}}=3$.} \label{radialfig}
\end{figure}

In our runs, we considered various lattice sizes (typically $16^4$) and $\beta$ values
in the range from 0.96 to 1.01. In the study of the $E_{x}$ profile, we chose interquark distances $d_{Q\bar{Q}}$ from 3 to 7; the results show rotational invariance and (at least qualitatively) the expected scaling properties as $\beta$ or $d_{Q\bar{Q}}$ are varied. The profile has a marked peak centered in the mid-point between the charges, and a fast (most likely exponential-like) decay as a function of $\rho$. Errorbars for the data shown in fig. \ref{radialfig}
are smaller than plotted symbols, but more precise statistics is needed in order to confirm or refute the large $\rho$ behaviour predicted by eq. (\ref{expectedex}).

\begin{figure}[htb]
\vspace{9pt}
\includegraphics[scale=0.44]{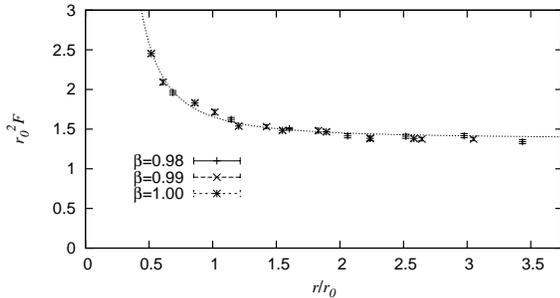}
\caption{Force as a function of the interquark distance, for various values of $\beta$.} \label{forzafig}
\end{figure}

As it concerns the interquark potential and force, in fig. \ref{forzafig}
we plot the interquark force $F$ \emph{versus} the $Q\bar{Q}$ distance $r$; $r_0$ is  Sommer's scale \cite{heplat9310022}. Notice the constant errorbars for different values of $r$. The dotted line is a fit to the theoretical expectation for $F(r)$ obtained by derivation of eq. (\ref{potential}). Our results are in agreement with \cite{heplat0311016}; data analysis for $V(r)$ and $F(r)$ shows that the string behaviour is indeed confirmed at large distances, whereas at shorter distances the role of possible further contributions beyond the L\"uscher term is not completely clear.

\section{CONCLUSIONS}
Our preliminary numerical results show qualitative agreement with the predictions for the electric flux induced by external charges. As it concerns the interquark potential and force, the effective string scenario appears to be confirmed at large interquark distances, whereas at shorter distances the picture breakdown seems not to be completely cured by including a boundary term. As it is suggested by a comparison among different gauge models \cite{heplat0207003,heplat0406008,heplat0403004},
we guess that a possible effective pattern at short distances might be \emph{non-universal}, \emph{i.e.} dependent on the gauge theory. Further details, larger statistics results, and
a more complete data analysis are published in ref.~\cite{Panero:2005iu}.

It is a pleasure to thank Michele Caselle for enlightening talks. The author acknowledges support received from Enterprise Ireland under the Basic Research Programme.

\end{document}